\documentclass[12pt]{article}
\begin{document}

\title{Quantum Information and \\ Spacetime Structure}
\date{~}
\author{Igor V. Volovich\\
 {\it Steklov Mathematical Institute, Russian Academy of Sciences,}\\
{\it Gubkin St.8,117966 Moscow, Russia}
\\E-mail: volovich@mi.ras.ru }

 \maketitle
\begin{abstract}
In  modern quantum information theory one deals with an idealized
situation  when the spacetime dependence of quantum phenomena is
neglected. However the transmission and processing of (quantum)
information is a physical process in spacetime. Therefore such
basic notions in quantum information theory as qubit, channel,
composite systems and entangled states should be formulated in
space and time.  In this paper some basic notions of quantum
information theory are considered from the point of view of
quantum field theory and general relativity. It is  pointed out an
important fact that in quantum field theory there is a statistical
dependence between two regions in spacetime even if they  are
spacelike separated. A classical probabilistic representation for
a family of correlation functions in quantum field theory is
obtained. A  noncommutative generalization of von Neumann`s
spectral theorem is discussed. We suggest a new physical principle
describing a relation between the mathematical formalism of
Hilbert space and quantum physical phenomena which goes beyond the
superselection rules. Entangled states and the change of state
associated with the measurement process in space and time are
discussed including the black hole and the cosmological spacetime.
 It is shown that any reasonable state in relativistic quantum
field theory becomes disentangled (factorizable) at large
spacelike distances if one makes local observations. As a result a
violation of Bell`s inequalities can be observed without
inconsistency with principles of relativistic quantum theory only
if the distance between detectors is rather small. We suggest a
further experimental study of entangled states in spacetime by
studying the dependence of the correlation functions on the
distance between detectors. \end{abstract}
\newpage
\section{Introduction}

Recent important  experimental and theoretical results obtained in
quantum computing, teleportation and cryptography (these topics
sometimes are considered as belonging to quantum information
theory) are based on the  investigation of  properties of
nonrelativistic quantum mechanics. Especially important are
properties of nonfactorized entangled states discussed by
Einstein, Podolsky and Rosen, Bohm, Bell and many others. Results
and ideas of Shannon`s classical information theory play an
important role in the modern quantum information theory as well as
the notions of qubit, quantum relative entropy,   quantum channel,
and entangled states , see for example ~\cite{QI} - \cite{Fuc}.

However the spacetime dependence is not explicitly indicated in
this approach. As a result, many important achievements  in modern
quantum information theory have been obtained for an idealized
situation when the spacetime dependence of quantum phenomena is
neglected.

We emphasize an importance of the investigation of quantum
information  in space and time. \footnote{The importance of the
investigation of quantum information effects in space and time and
especially the role of relativistic invariance in classical and
quantum information theory was stressed in the talk by the author
at the First International Conference on Quantum Information which
was held at Meijo University, Japan, November 4-8, 1997.}
Transmission and processing of (quantum) information is a physical
process in spacetime.  Therefore a formulation of such basic
notions in quantum information theory as  composite systems,
entangled states and the channel should include the spacetime
variables \cite{Vol3}.

Ultimately, quantum information theory should become
 a part  of quantum field theory (perhaps, in future, a part
 of superstring theory) since quantum field theory is
 our most fundamental physical theory.

Quantum field theory \cite{BSch} is not just an abstract
mathematical theory of operators in a Hilbert space. Basic
equations of quantum field theory such as the Maxwell, Dirac,
Yang--Mills equations are differential equations for operator
functions defined on the spacetime. The nonrelativistic
Schrodinger equation is also a differential equation in spacetime.
Therefore a realistic quantum information theory should be based
on the study of the solutions of these equations propagated in
spacetime including the curved spacetime.

Entangled states, i.e. the states of two particles with the wave
function which is not a product of the wave functions of single
particles,  have been studied in many theoretical and experimental
works starting from the paper of Einstein, Podolsky and Rosen, see
e.g.~\cite{AfrSel}.

In this paper entangled states in space and time are considered.
We point out a simple but the fundamental fact that the vacuum
state $\omega_0$ in a free quantum field theory is a nonfactorized
(entangled) state for observables belonging to spacelike separated
regions:
$$
\omega_0 (\varphi (x)\varphi (y)) - \omega_0 (\varphi (x))
\omega_0 (\varphi (y)) \neq 0$$ Here $\varphi (x) $ is a free
scalar field in the Minkowski spacetime and $ (x-y)^2 < 0$. Hence
there is a statistical dependence between causally disconnected
regions.

However one has an asymptotic factorization of the vacuum state
for  large separations of the spacelike regions. Moreover one
proves that in quantum field theory {\it there is an asymptotic
factorization for any reasonable state and any local observables.
Therefore at large distances any reasonable state becomes
disentangled}. We have the relation
$$
\lim_{|l|\to\infty}[\omega(A(l)B)-\omega(A(l))\omega(B)]=0
$$
Here $\omega$ is a state from a rather wide class of the states
which includes entangled states, $A$ and $B$ are two local
observables, and $A(l)$ is the translation of the observable $A$
along the 3 dim vector $l$. As a result a violation of Bell`s
inequalities (see below) can be observed without inconsistency
with principles of relativistic quantum theory only if the
distance between detectors is rather small. We suggest a further
experimental study of entangled states in spacetime by studying
the dependence of the correlation functions on the distance
between detectors.

There is no a factorization of the expectation value $\omega_0
(\varphi (x)\varphi (y)) $ even for the space-like separation of
the variables $x$ and $y$ if the distance between $x$ and $y$ is
not large enough. However we will prove that there exist a
representation of the form
$$
\omega_0 (\varphi (x)\varphi (y)) =E\xi (x)\xi^*(y)$$ which is
valid for all $x$ and $y$. Here $\xi (x)$ is a classical
(generalized) complex  random field and $E$ is the
 expectation value. Therefore the quantum correlation function
is represented as a classical correlation function of separated
random fields. This representation can be called a local realistic
representation by analogy with the Bell approach to the spin
correlation functions.

J. Bell proved ~\cite{Bel} that  there are quantum spin
correlation functions in entangled states that can not be
represented as classical correlation functions of separated random
variables. Bell's theorem reads, see \cite{Vol1}:
$$
\cos(\alpha - \beta)\neq E\xi_{\alpha}\eta_{\beta}
$$
where $\xi_{\alpha}$ and $\eta_{\beta}$ are two random processes
such that $|\xi_{\alpha}|\leq 1$,~ $|\eta_{\beta}|\leq 1$ and $E$
is the expectation. Here the function $\cos(\alpha - \beta)$
describes the quantum mechanical correlation of spins of two
entangled particles. Bell`s theorem has been interpreted as
incompatibility of the requirement of locality with the
statistical predictions of quantum mechanics~\cite{Bel}. For a
recent discussion of Bell's theorem and Bell`s inequalities see,
for example ~\cite{AfrSel} - ~\cite{VV2} and references therein.

However if we want to speak about locality in quantum theory then
we have to localize somehow our particles. For example we could
measure the density of the energy or the position of the particles
simultaneously with the spin. Only then we could come to some
conclusions about a relevance of the spin correlation function to
the problem of locality.

 The function
$\cos(\alpha - \beta)$ describes quantum correlations of two spins
in the two qubit Hilbert space when the spacetime dependence of
the wave functions of the particles is neglected. Let us note
however that the very formulation of the problem of locality in
quantum mechanics prescribes a special role to the position in
ordinary three-dimensional space. It is rather strange therefore
that the problem of {\it local in space observations} was
neglected in   discussions of the problem of locality in relation
to Bell's inequalities .

Let us stress  that we discuss here not a problem of
interpretation of quantum theory but a problem of how to make
correct quantum mechanical computations describing an experiment
with two  detectors localized in space. Recently it was  pointed
out \cite{Vol1} that  if we make  {\it local} observations of
spins then the spacetime part of the wave function leads to an
extra factor in quantum correlations and as a result the ordinary
conclusion from the Bell theorem about the nonlocality of quantum
theory fails.

 We present a modification of Bell`s equation which
includes space and time variables. The function $\cos(\alpha -
\beta)$ describes the quantum mechanical correlation of spins of
two entangled particles if we neglect the spacetime dependence of
the wave function. It was shown in \cite{Vol1} that if one takes
into account the space part of the wave function then  the quantum
correlation describing local observations of spins in the simplest
case will take the form $g \cos (\alpha - \beta)$ instead of just
$\cos (\alpha - \beta)$. Here the parameter $g=g ({\cal O}_A,{\cal
O}_B)$ describes the location of the detectors in regions ${\cal
O}_A$ and ${\cal O}_B$ in space. In this case we have to
investigate a modified Bell`s equation. We will prove that there
exists the following representation
$$
g ({\cal O}_A,{\cal O}_B)\cos(\alpha-\beta)
 =E\xi (\alpha,{\cal O}_A)\eta (\beta,{\cal O}_B)
 $$
if the distance between ${\cal O}_A$ and ${\cal O}_B$ is large
enough. We will show that in fact at large distances all
reasonable quantum states become disentangled.
 This fact leads also to important consequences for
quantum teleportation and quantum cryptography,  \cite{Vol4,VV2}.
Bell's theorem constitutes an important part in quantum
cryptography. In \cite{Vol4} it is discussed how one can try to
improve the security of quantum cryptography schemes in space  by
using  a special preparation of the space part of the wave
function.

It is important  to study also a more general question: which
class of functions $f(s,t)$ admits a representation of the form
$$
f(s,t)=Ex_sy_t
$$
where $x_s$ and $y_t$ are  bounded stochastic processes and also
analogous question for the functions of several variables
$f(t_1,...,t_n).$

Such considerations could provide a {\it noncommutative}
generalization of von Neumann`s spectral theorem. We suggest a new
physical principle describing a relation between the mathematical
formalism of Hilbert space and quantum physical phenomena.

In modern quantum information theory the basic notion is the two
dimensional Hilbert space, i.e. qubit. We suggest that in a
relativistic quantum information theory, when the existence of
spacetime is taken into account, the basic notion should be a
notion of an elementary quantum system, i.e. according to Wigner
(see \cite{Bog}) it is an infinite dimensional Hilbert space $H$
invariant under an irreducible representation of the Poincare
group labelled by $[m,s]$ where $m\geq 0$ is  mass and
$s=0,1/2,1,...$ is  spin (helicity).

In the next section the disentanglement at large distances in
quantum field theory is considered. Local observations and
modified Bell`s equations are considered in Sect.3. Noncommutative
spectral theory and local realism are discussed in Sect.4. Some
remarks on the properties of entangled states in curved spacetime
are made in Sect.5.

\section{Quantum Probability and Quantum Field Theory}

In quantum probability (see \cite{ALV}) we are given a * - algebra
${\cal A}$ and a state (i.e. a linear positive normalized
functional) $\omega$ on ${\cal A}$. Elements from ${\cal A}$ are
called random variables. Two random variables $A$ and $B$ are
called (statistically) independent if $
\omega(AB)=\omega(A)\omega(B)$.

Quantum field in the Wightman formulation is an operator-valued
distribution $\varphi (f)$ acting in a Hilbert space where $f$ is
a Schwartz test function on $R^4$. One uses the standard notations
$$
\varphi (f) =\int_{R^4}\varphi (x)f(x)dx
$$
Quantum field satisfies simple transformation properties under a
representation of the Poincare group. Moreover if $f$ and $g$ are
test functions whose supports are space-like to each other then
$\varphi (f)$ and $\varphi (g)$ shall commute:
$$
[\varphi (f),\varphi (g)]=0
$$
  This assumption rests
on the principle that no physical effect can propagate in
space-like directions. The assumption is called the microscopic or
relativistic causality.

We point out an interesting fact that  for the vacuum state in
quantum field theory the relativistic causality does not lead to
the statistical independence in the sense of quantum probability
for quantum fields with weight functions whose supports are
space-like to each other. We will prove the following

{\bf Proposition 1}. {\it There is a statistical dependence
between two spacelike separated regions for the vacuum state in
the theory of free scalar quantum field}.

{\bf Proof}. Let us consider a free scalar quantum field $\varphi
(x)$:
$$
\varphi (x) =\frac{1}{(2\pi)^{3/2}}\int_{R^3} \frac{d{\bf
k}}{\sqrt{2k^0}}(e^{ikx}a^*({\bf k})+ e^{-ikx}a({\bf k}))
$$
Here $kx=k^0x^0-{\bf kx}$, $k^0=\sqrt{{\bf k}^2+m^2}, m\geq 0$ and
$a({\bf k})$ and $a^*({\bf k})$ are  annihilation and creation
operators,
$$
[a({\bf k}), a^*({\bf k'})]=\delta ({\bf k}-{\bf k'})
$$
The field $\varphi (x)$ is an operator valued distribution acting
in the Fock space ${\cal F}$ with the vacuum $|0>$,
$$
a({\bf k})|0>=0
$$
The vacuum expectation value of two fields is
$$
\omega_0 (\varphi (x)\varphi (y))=<0|\varphi (x)\varphi (y)|0>
=W_0(x-y, m^2)
$$
where
$$
W_0(x-y, m^2)=\frac{1}{(2\pi)^{3}}\int_{R^3} \frac{d{\bf
k}}{2k^0}e^{-ik(x-y)}
$$
The statistical independence of two spacelike separated regions in
particular would lead to the relation
$$
\omega_0 (\varphi (x)\varphi (y)) - \omega_0 (\varphi (x))
\omega_0 (\varphi (y)) = 0
$$
if $(x-y)^2<0$. But since $\omega_0 (\varphi (x))=0$ in fact we
have
$$
\omega_0 (\varphi (x)\varphi (y)) - \omega_0 (\varphi (x))
\omega_0 (\varphi (y)) = W_0(x-y, m^2)\neq 0
$$
Therefore it is proved that there is a statistical dependence
between the spacelike separated regions for the vacuum state in
theory of free scalar quantum field. The proposition is
proved.

Note however that the violation of the statistical independence
vanish exponentially with the spacial separation of $x$ and $y$
since for large $\lambda =m\sqrt {-x^2}$ the function $W_0(x,
m^2)$ behaves like
$$
\frac{m^2}{4\pi \lambda}\left(\frac{\pi}
{2\lambda}\right)^{1/2}e^{-\lambda}
$$

Let us prove that any polynomial state is asymptotically
disentangled (factorized) for large spacelike distances. Let
${\cal A}$ be the algebra of polynomials in the Fock space ${\cal
F}$ at the field $\varphi (f)$ with the test functions $f$ . Let
$C\in {\cal A}$ and $|\psi>=C|0>.$ Denote the state $\omega
(A)=\left<\psi|A|\psi\right >/||\psi||^2$ for $A\in {\cal A}$.

{\bf Theorem 2.} One has the following asymptotic disentanglement
property
$$
\lim_{|l|\to\infty}[\omega(A(l)B)-\omega(A(l))\omega(B)]=0
$$
Here  $A$ and $B$ belong to ${\cal A}$ and $A(l)$ is the
translation of $A$ along the 3 dim vector $l$. One has also
$$
\lim_{|l|\to\infty}[\omega(A(l))-\left <0|A(l)|0\right >]=0
$$
The proof of the theorem is based on the Wick theorem and the
Riemann-Lebesgue lemma.

Similar theorems take place also for the Dirac and the Maxwell
fields. In particular for the Dirac field $\psi (x)$ one can prove
the asymptotic factorization for the local spin operator
$$
{\bf S}({\cal O})=\int_{{\cal O}}\psi^*{\bf \Sigma}\psi dx
$$
Here $\Sigma$ is made from the Dirac matrices.

Finally let us show that some correlation functions in the
relativistic quantum field theory can be represented as
mathematical expectations of the classical (generalized) random
fields.

{\bf Theorem 3.} If $\varphi (x)$ is a scalar complex quantum
field then one has a representation
$$
\left <0|\varphi (x_1)...\varphi (x_n)\varphi^* (y_1) ...\varphi^*
(y_n)|0\right >=E\xi (x_1)...\xi (x_n)\xi^* (y_1) ...\xi^* (y_n).
$$
Here $\xi (x)$ is a complex random field.

The proof of the theorem follows from the positivity of the
quantum correlation functions. It is interesting that we have
obtained a functional integral representation for the quantum
correlation functions in real time. Similar representation is
valid also for the 2-point correlation function of an interacting
scalar field. It follows from the Kallen-Lehmann representation.

\section{Local Observations and Modified Bell's Equations}

 Bell's theorem reads:
\begin{equation}
\cos(\alpha - \beta)\neq E\xi_{\alpha}\eta_{\beta} \label{eq:cos}
\end{equation}
where $\xi_{\alpha}$ and $\eta_{\beta}$ are two random processes
such that $|\xi_{\alpha}|\leq 1$,~ $|\eta_{\beta}|\leq 1$ and $E$
is the expectation. In more details:

{\bf Theorem 4.} There exists no probability space $(\Lambda,
{\cal F}, d\rho (\lambda))$ and a pair of stochastic processes
$\xi_{\alpha}=\xi_{\alpha}(\lambda),
~\eta_{\beta}=\eta_{\beta}(\lambda), ~0\leq \alpha,\beta \leq
2\pi$ which obey $|\xi_{\alpha}(\lambda)|\leq 1$,~
$|\eta_{\beta}(\lambda)|\leq 1$ such that the following equation
is valid
\begin{equation}
\cos(\alpha - \beta) = E\xi_{\alpha}\eta_{\beta} \label{eq:cosin}
\end{equation}
for all $\alpha$ and $\beta$.

Here $\Lambda$ is a set, ${\cal F}$ is a sigma-algebra of subsets
and $d\rho (\lambda)$ is a probability measure, i.e. $d\rho
(\lambda) \geq 0,~\int d\rho (\lambda)=1.$ The expectation is
$$
E\xi_{\alpha}\eta_{\beta}=\int_{\Lambda}\xi_{\alpha}(\lambda)
\eta_{\beta}(\lambda)d\rho (\lambda)
$$

The theorem follows from the CHSH inequality presented below. Let
us discuss  a physical interpretation of this result.

Consider a pair of spin one-half particles formed in the singlet
spin state and moving freely towards two detectors.  If one
neglects the space part of the wave function  then one has the
Hilbert space $C^2\otimes C^2$ and  the quantum mechanical
correlation of two spins in the singlet state $\psi_{spin}\in
C^2\otimes C^2$ is
\begin{equation}
 D_{spin}(a,b)=\left<\psi_{spin}|\sigma\cdot a \otimes\sigma\cdot
b|\psi_{spin}\right>=-a\cdot b \label{eq:eqn1}
\end{equation}
Here $a=(a_1,a_2,a_3)$ and $b=(b_1,b_2,b_3)$ are two unit vectors
in three-dimensional space $R^3$,
$\sigma=(\sigma_1,\sigma_2,\sigma_3)$ are the Pauli matrices,
$$
\sigma_1=\left(
    \begin{array}{cc}0 & 1\\ 1 & 0
    \end{array}
\right),~~ \sigma_2=\left(
    \begin{array}{cc}0 & -i\\ i & 0
    \end{array}
\right),~~ \sigma_3=\left(
    \begin{array}{cc}1 & 0\\ 0 & -1
    \end{array}
\right),~~
 \sigma\cdot a =\sum_{i=1}^{3}\sigma_i a_i
 $$
 and
$$
\psi_{spin}=\frac{1}{\sqrt 2} \left(\left(
    \begin{array}{c}0\\1
    \end{array}
    \right)
\otimes \left(
    \begin{array}{c}1\\
    0\end{array}
    \right)
-\left(
    \begin{array}{c}1\\
    0\end{array}
    \right)
\otimes \left(
    \begin{array}{c}0\\
    1\end{array}
    \right)
\right)
$$
If the vectors $a$ and $b$ belong to the same plane then one can
write $-a\cdot b=\cos (\alpha - \beta)$ and hence Bell's theorem
states that the function $ D_{spin}(a,b)$ Eq.~(\ref{eq:eqn1}) can
not be represented in the form
\begin{equation}
\label{eq:eqn2} P(a,b)=\int \xi (a,\lambda) \eta (b,\lambda)
d\rho(\lambda)
\end{equation}
i.e.
\begin{equation}
\label{eq:Ab} D_{spin}(a,b)\neq P(a,b)
\end{equation}
Here $ \xi (a,\lambda)$ and $  \eta(b,\lambda)$ are random  fields
on the sphere, $|\xi (a,\lambda)|\leq 1$,~  $ | \eta
(b,\lambda)|\leq 1$ and $d\rho(\lambda)$ is a positive probability
measure,  $ \int d\rho(\lambda)=1$. The parameters $\lambda$ are
interpreted as hidden variables in a realist theory. It is clear
that Eq.~(\ref{eq:Ab}) can be reduced to Eq.~(\ref{eq:cos}).

To prove Theorem 4 one uses the following theorem which is a
slightly generalized  Clauser-Horn-Shimony-Holt (CHSH)  result.

{\bf Theorem 5.} Let $f_1,~f_2,~g_1$ and $g_2$ be random variables
(i.e. measured functions) on the probability space $(\Lambda,
{\cal F}, d\rho (\lambda))$ such that
$$
|f_i(\lambda)g_j(\lambda)|\leq 1,~~i,j=1,2.
$$
Denote
$$
P_{ij}=Ef_ig_j,~~i,j=1,2.
$$
Then
$$
|P_{11}-P_{12}|+|P_{21}+P_{22}|\leq 2.
$$
The last inequality is called
  the CHSH inequality.
By using notations of Eq.~(\ref{eq:eqn2}) one has
\begin{equation}
\label{eq:eqn3}
 |P(a,b)-P(a,b')|+|P(a',b)+P(a',b')|\leq 2
\end{equation}
for any four unit vectors $a,b,a',b'$.

It will be shown below that if one takes into account the space
part of the wave function then  the quantum correlation in the
simplest case will take the form $g \cos (\alpha - \beta)$ instead
of just  $\cos (\alpha - \beta)$ where the parameter $g$ describes
the location of the system in space and time. In this case one can
get a representation
\begin{equation}
g\cos(\alpha - \beta)= E\xi_{\alpha}\eta_{\beta} \label{eq:gcos}
\end{equation}
if $g$ is small enough. The factor $g$ gives a contribution to
visibility or efficiency of detectors that are used in the
phenomenological description of detectors.

\subsection{Modified Bell`s equation}

In the previous section the space part of the wave function of the
particles was neglected. However exactly the space part is
relevant to the discussion of locality. The Hilbert space assigned
to one particle with spin 1/2 is  $C^2\otimes L^2(R^3)$ and the
Hilbert space of two particles is $C^2\otimes L^2(R^3)\otimes
C^2\otimes L^2(R^3).$ The complete wave function is $\psi
=(\psi_{\alpha\beta}({\bf r}_1,{\bf r}_2,t))$ where $\alpha$ and
$\beta $ are spinor indices, $t$ is time  and ${\bf r}_1$ and
${\bf r}_2$ are vectors in three-dimensional space.

We suppose that there are two  detectors (A and B) which are
located in space $R^3$ within the two localized regions ${\cal
O}_A$ and ${\cal O}_B$ respectively, well separated from one
another. If one makes a local observation in the region ${\cal
O}_A$ then this means that one measures not only the spin
observable $\sigma_i$ but also some another observable which
describes the localization of the particle like the energy density
or the projection operator $P_{{\cal O}}$ to the region ${\cal
O}$. We will consider here correlation functions of the projection
operators $P_{{\cal O}}$.

Quantum correlation describing the localized measurements of spins
in the regions ${\cal O}_A$ and ${\cal O}_B$ is

\begin{equation}
\label{eq:eqn6} \omega(\sigma\cdot a   P_{{\cal O}_A}\otimes
\sigma\cdot b  P_{{\cal O}_B})=\left<\psi| \sigma\cdot a
P_{{\cal O}_A}\otimes  \sigma\cdot b  P_{{\cal O}_B} |\psi\right>
\end{equation}

Let us consider the simplest case when the wave function has the
form of the product of the spin function and the space function
$\psi=\psi_{spin}\phi({\bf r}_1,{\bf r}_2)$. Then one has
\begin{equation}
\label{eq:eqn7}
 \omega(\sigma\cdot a   P_{{\cal O}_A}\otimes
\sigma\cdot b  P_{{\cal O}_B})=
 =g ({\cal O}_A,{\cal O}_B)
  D_{spin}(a,b)
\end{equation}
where the function
\begin{equation}
\label{eq:eqn8}
 g ({\cal O}_A,{\cal O}_B)=\int_{{\cal O}_A \times {\cal O}_B}|\phi({\bf
 r}_1,{\bf
 r}_2)|^2 d{\bf r}_1d{\bf r}_2
\end{equation}
describes correlation of particles in space. It is the probability
to find one particle in the region ${\cal O}_A$ and another
particle in the region ${\cal O}_B$.

One has
\begin{equation}
\label{eq:eqn8g} 0\leq g ({\cal O}_A,{\cal O}_B)\leq 1.
\end{equation}

If ${\cal O}_A$ is a bounded region and ${\cal O}_A(l)$ is a
translation of ${\cal O}_A$ to the 3-vector $l$ then one has

\begin{equation}
\label{eq:eqn8l} \lim_{|l|\to\infty} g({\cal O}_A(l),{\cal
O}_B)=0.
\end{equation}

Since
$$\left<\psi_{spin}|\sigma\cdot a \otimes
I|\psi_{spin}\right>=0
$$
we have
$$
\omega (\sigma\cdot a P_{{\cal O}_A}\otimes I)=0.
$$
Therefore we have proved the following proposition which says that
the state  $\psi=\psi_{spin}\phi({\bf r}_1,{\bf r}_2)$ becomes
disentangled at large distances.

{\bf Proposition 6.} One has the following property of the
asymptotic factorization (disentanglement) at large distances:
 \begin{equation}
\label{eq:eqn8ld} \lim_{|l|\to\infty} [\omega (\sigma\cdot a
P_{{\cal O}_A(l)}\otimes \sigma\cdot b P_{{\cal O}_B})- \omega
(\sigma\cdot a P_{{\cal O}_A(l)}\otimes I )\omega(I\otimes
\sigma\cdot b P_{{\cal O}_B} )]=0
\end{equation}
or
$$
\lim_{|l|\to\infty} \omega (\sigma\cdot a P_{{\cal O}_A(l)}\otimes
\sigma\cdot b P_{{\cal O}_B})=0.
$$
Now one inquires whether one can write a representation
\begin{equation}
\label{eq:eqn9}
 \omega(\sigma\cdot a   P_{{\cal O}_A(l)}\otimes
\sigma\cdot b  P_{{\cal O}_B})=
 \int \xi (a,{\cal O}_A,\lambda)
 \eta (b,{\cal O}_B,\lambda) d\rho(\lambda)
\end{equation}
where $|\xi (a,{\cal O}_A(l),\lambda)|\leq 1,~~ |\eta (b,{\cal
O}_B,\lambda)|\leq 1$.

{\bf Remark.} A local modified equation reads
$$
|\phi ({\bf r_1},{\bf r_2},t)|^2\cos(\alpha - \beta) =E\xi
(\alpha,{\bf r_1},t) \eta (\beta,{\bf r_2},t).
$$

If we are interested in the conditional probability of finding the
projection of spin along  vector $a$ for the particle 1  in the
region ${\cal O}_A(l)$ and the projection of spin along the vector
$b$ for the particle 2 in the region ${\cal O}_B$   then we have
to divide both sides of Eq.~(\ref{eq:eqn9}) by $g({\cal
O}_A(l),{\cal O}_B)$.

Note  that here the classical random variable $\xi=\xi (a,{\cal
O}_A(l),\lambda)$ is not only separated in the sense of Bell (i.e.
it depends only on $a$) but it is also local in the 3 dim space
since it depends only on the region ${\cal O}_A(l)$. The classical
random variable $\eta$ is also local in 3 dim space since it
depends only on ${\cal O}_B$. Note also that since the eigenvalues
of the projector $P_{{\cal O}}$ are 0 or 1 then one should have $
|\xi (a,{\cal O}_A)|\leq 1.$

Due to the property of the asymptotic factorization and the
vanishing of the quantum correlation for large $|l|$ there exists
a trivial asymptotic classical representation of the form
(\ref{eq:eqn9}) with $\xi=\eta=0.$

We can do even better and find a classical representation which
will be valid uniformly for large $|l|$.

If $g$ would not depend on ${\cal O}_A$ and ${\cal O}_B$ then
instead of Eq~(\ref{eq:cosin}) in Theorem 1  we could have a
modified equation
\begin{equation}
g\cos(\alpha - \beta) = E\xi_{\alpha}\eta_{\beta}
\label{eq:cosinus}
\end{equation}
The factor $g$ is important. In particular one can write the
following representation \cite{VV} for $0\leq g\leq 1/2$:
\begin{equation}
\label{eq:gek} g\cos(\alpha-\beta)= \int_0^{2\pi}\sqrt
{2g}\cos(\alpha-\lambda) \sqrt {2g}\cos(\beta-\lambda)
 \frac{d\lambda}{2\pi}
\end{equation}
Therefore if $0\leq g\leq 1/2$ then there exists a solution of
Eq.~(\ref{eq:cosinus}) where
$$
\xi_{\alpha}(\lambda)=\sqrt {2g}\cos(\alpha-\lambda),~
~\eta_{\beta}(\lambda)=\sqrt {2g}\cos(\beta-\lambda)
$$
and $|\xi_{\alpha}|\leq 1,~|\eta_{\beta}|\leq 1.$ If $g>1/\sqrt 2$
then it follows from Theorem 2 that there is no solution to
Eq.~(\ref{eq:cosinus}). We have obtained

{\bf Theorem 7.} If $g>1/\sqrt 2$ then there is no solution
$(\Lambda, {\cal F}, d\rho (\lambda), \xi_{\alpha}, \eta_{\beta})$
to Eq.~(\ref{eq:cosinus}) with the bounds $|\xi_{\alpha}|\leq 1$,~
$|\eta_{\beta}|\leq 1.$ If $0\leq g\leq 1/2$ then there exists a
solution to Eq.~(\ref{eq:cosinus}) with the bounds
$|\xi_{\alpha}|\leq 1$,~ $|\eta_{\beta}|\leq 1.$

{\bf Remark.} Local variable models for inefficient detectors are
presented in \cite{San,Lar}.

Let us take now the wave function $\phi$ of the form
$\phi=\psi_{1}({\bf r}_1)\psi_{2}({\bf r}_2)$  where
$$
\int_{R^3}|\psi_{1}({\bf r}_1)|^2d{\bf r}_1=1,~~
\int_{R^3}|\psi_{2}({\bf r}_2)|^2d{\bf r}_2=1
$$
In this case
$$
 g ({\cal O}_A(l),{\cal O}_B)=
 \int_{{\cal O}_A(l)}|\psi_{1}({\bf r}_1)|^2d{\bf r}_1\cdot
 \int_{{\cal O}_B}|\psi_{2}({\bf r}_2)|^2d{\bf r}_2
$$
There exists such $L>0$ that
$$
\int_{B_L}|\psi_{1}({\bf r}_1)|^2d{\bf r}_1=\epsilon <1/2,~~
$$
where $B_L=\{{\bf r}\in R^3: |{\bf r}|\geq L\}.$ Let us  make an
additional assumption that the classical random variable  has the
form of a product of two independent classical random variables $
\xi (a,{\cal O}_A)=\xi_{space}({\cal O}_A)\xi_{spin}(a)$ and
similarly for $\eta.$ We will prove that there exists the
following representation
$$
g ({\cal O}_A,{\cal O}_B)\cos(\alpha-\beta)
 =E\xi (\alpha,{\cal O}_A)\eta (\beta,{\cal O}_B)
 $$
if the distance between ${\cal O}_A$ and ${\cal O}_B$ is large
enough. We have the following

{\bf Theorem 8.} Under the above assumptions and for large enough
$|l|$ there exists the following representation of the quantum
correlation function
$$
 g ({\cal O}_A(l),{\cal O}_B)\cos(\alpha-\beta)
 =(E\xi_{space}({\cal O}_A)(l))(E\eta_{space}({\cal O}_B))
 E\xi_{spin}(\alpha)\xi_{spin}(\beta)
 $$
 where all classical random variables are bounded by 1.

{\bf Proof.} To prove the theorem we write
$$
g ({\cal O}_A(l),{\cal O}_B)\cos(\alpha-\beta) =\int_{{\cal
O}_A(l)}\frac{1}{\epsilon}|\psi_{1}({\bf r}_1)|^2d{\bf r}_1 \cdot
 \int_{{\cal O}_B}|\psi_{2}({\bf r}_2)|^2d{\bf r}_2
\cdot \epsilon \cos (\alpha-\beta)
$$
$$=(E\xi_{space}({\cal O}_A(l))(E\eta_{space}({\cal O}_B))
 E\xi_{spin}(\alpha)\xi_{spin}(\beta)
$$
Here  $\xi_{space}({\cal O}_A(l))$ and $\eta_{space}({\cal O}_B)$
are random variables on the probability space $B_L\times R^3$ with
the probability measure
$$
dP({\bf r}_1,{\bf r}_2)= \frac{1}{\epsilon}|\psi_{1}({\bf r}_1)|^2
\cdot
 |\psi_{2}({\bf r}_2)|^2d{\bf r}_1d{\bf r}_2
$$
of the form
$$
\xi_{space}({\cal O}_A(l),{\bf r}_1,{\bf r}_2) =\chi_{{\cal
O}_A(l)}({\bf r}_1),~~ \eta_{space}({\cal O}_B,{\bf r}_1,{\bf
r}_2) =\chi_{{\cal O}_B}({\bf r}_2)$$ where $\chi_{{\cal O}}({\bf
r})$ is the characteristic function of the region ${\cal O}.$ We
assume that ${\cal O}_A(l)$ belongs to $B_L.$ Further
$\xi_{spin}(\alpha)$ is a random process on the circle $0\leq
\varphi \leq 2\pi$ with the probability measure $d\varphi /2\pi$
of the form
$$
\xi_{spin}(\alpha,\varphi)=\sqrt{2\epsilon}\cos(\alpha - \varphi)
$$
The theorem is proved.

\section{Noncommutative Spectral Theory and Quantum Theory}

As a generalization of the previous discussion we would like to
suggest here a new general physical principle which describes a
relation between the mathematical formalism of Hilbert space and
physical quantum phenomena. It will use  theory of classical
stochastic processes \cite{Hid}
 which, as we suggest, expresses the condition of local realism.
 According to the standard view to quantum theory any hermitian
 operator in a Hilbert space describes a physical observable
 and any density operator describes a physical state. Here we
 would like to suggest that this view is too general and that
 in fact there should exist some additional restrictions to the
 family of Hermitian operators and to the density operators if they
 have to describe physical phenomena.

Let ${\cal H}$ be a Hilbert space, $\rho $ is the density
operator, $\{A_{\alpha}\}$ is a family of self-adjoint operators
in ${\cal H}$. One says that the family of observables
$\{A_{\alpha}\}$ and the state $ \rho$ satisfy to {\it the
condition of local realism} if there exists a probability space
$(\Lambda, {\cal F}, dP (\lambda))$ and a family of random
variables $\{\xi_{\alpha}\}$ such that  the range of
$\xi_{\alpha}$ belongs to the spectrum of $A_{\alpha}$ and for any
subset $\{ A_{i_1},...,A_{i_n}\}$ of mutually commutative
operators one has a representation
$$
Tr( \rho A_{i_1}...A_{i_n})=E\xi_{i_1}...\xi_{i_n}
$$
The physical meaning of the representation is that it describes
the quantum-classical correspondence. If the family
$\{A_{\alpha}\}$ would be a maximal commutative family of
self-adjoint operators then for pure states the previous
representation can be reduced to the von Neumann spectral
theorem~\cite{Nai}. In our case the family $\{A_{\alpha}\}$
consists  from not necessary commuting operators. Hence we will
call such a representation a {\it noncommutative spectral
representation}.  Of course one has a question for which families
of operators and states a {\it noncommutative spectral theorem}
is valid, i.e. when we can write the noncommutative spectral
representation. We need a noncommutative generalization of von
Neumann`s spectral theorem.

It would be helpful to study the following problem: describe the
class of functions $f(t_1,...,t_n)$ which admits the
representation of the form
$$
f(t_1,...,t_n)=Ex_{t_1}...z_{t_n}
$$
where $x_t,...,z_t$ are random processes which obey the bounds
$|x_t|\leq 1,...,|z_t|\leq 1$.

From the previous discussion (Bell`s theorem) we know that there
are such families of operators and such states which do not admit
the noncommutative spectral representation and therefore they do
not satisfy the condition of local realism. Indeed let us take the
Hilbert space ${\cal H}=C^2\otimes C^2$ and  operators
$\sigma\cdot a \otimes  \sigma\cdot b$. We know from Theorem 2
that the function $\left<\psi_{spin}|\sigma\cdot a \otimes
\sigma\cdot b|\psi_{spin}\right>$ can not be represented as the
expected value $E\xi (a)\eta (b)$  of random variables.

However, as it was discussed above, the space part of the wave
function was neglected in the previous consideration. It was
proved that for the observables of the form $\sigma\cdot a
P_{{\cal O}_A}\otimes  \sigma\cdot b  P_{{\cal O}_B}$ one can
write a local spectral representation if the distance between the
regions ${\cal O}_A$ and ${\cal O}_B$ is large enough. We suggest
that {\it in physics one could prepare only such states and
observables which satisfy the condition of local realism.} Perhaps
we should restrict ourself in this proposal to the consideration
of only such families of observables which satisfy the condition
of relativistic local causality. If there are physical phenomena
which do not satisfy this proposal then it would be important to
{\it describe quantum processes which satisfy the above formulated
condition of local realism and also processes which do not satisfy
to this condition}.

\section{Entangled States and General Relativity}
Here we would like to make some comments on the study of entangled
states and the reduction postulate in curved spacetime. It is
especially interesting to consider  properties of entangled states
in curved spacetimes possessing a nontrivial causal structure in
particular in a spacetime containing an event horizon. In
particular entangled states and the reduction of the wave function
in the context of  black holes, Hawking radiation, inflationary
models of the early universe, creation of particles and
accelerated detectors \cite{FN}-\cite{BD} should be considered.
Inflation leads to the phenomena of the cosmic entanglement since
the scalar field creates particles during the inflation
\cite{Lin}.Analysis of quantum teleportation of a state through
the horizon can help to clarify the notion of the reduction of the
wave function associated with the measurement process.

\section {Conclusions}

We have discussed some problems in quantum information theory
which requires the inclusion of spacetime variables. In particular
entangled states in space and time were considered. A modification
of Bell`s equation which includes the spacetime variables is
investigated. A general relation between quantum theory and theory
of classical stochastic processes was proposed
 which expresses the condition of local realism
 in the form of a noncommutative spectral theorem.
 Entangled states   in space and time are
considered. It is  shown that any reasonable state in relativistic
quantum field theory  becomes disentangled (factorizable) at large
space-like distances if one makes local observations. As a result
a  violation of Bell`s  inequalities can be observed without
inconsistency with principles of relativistic quantum theory only
if the distance between detectors is rather small.

 There are many
interesting open problems in the approach to quantum information
in space and time discussed in this paper. Some of them related
with the noncommutative spectral theory and theory of classical
stochastic processes.
 We suggest a
further experimental study of entangled states in spacetime by
studying the dependence of the correlation functions on the
distance between detectors. It is very interesting to investigate
properties of entangled states in curved spacetime.

\section*{Acknowledgments}
I am grateful to G.G. Emch, R. Gill, Y.S. Kim, B. Hiley, G.` t
Hooft, A. Khrennikov, W. Philipp, H. Rauch, and A. Sadreev for
useful discussions. This work is supported in part by
RFFI-02-01-01084 and the grant for the leading scientific schools
00-15-96073.

\end{document}